\documentclass[twocolumn,preprintnumbers,secnumarabic,amsmath,amssymb,superscriptaddress,nofootinbib]{revtex4}

\usepackage{graphicx}
\usepackage{dcolumn}
\usepackage{bm}
\usepackage{paralist}

\def\be{\begin{equation}}
\def\ee{\end{equation}}
\def\bea{\begin{eqnarray}}
\def\eea{\end{eqnarray}}

\begin{document}

\title{$\overline{D3}$'s -- Singular to the Bitter End}
\smallskip
\author{Iosif Bena, Mariana Gra\~na, Stanislav Kuperstein and Stefano Massai}

\affiliation{{Institut de Physique Th\'eorique, CEA
Saclay, CNRS URA 2306, F-91191 Gif-sur-Yvette, France \\
  \textsf{iosif.bena, mariana.grana, stanislav.kuperstein, stefano.massai@cea.fr}
}}

\begin{abstract}

We study the full backreaction of anti-D3 branes smeared over the tip of the deformed conifold.
Requiring the 5-form flux and warp factor at the tip to be that of anti-D3 branes, we find a simple
power counting argument showing that if the three-form fluxes have no IR singularity, they will be necessarily imaginary-anti-self-dual. Hence the only solution with anti-D3 branes at the tip of the conifold that is regular in the IR and the UV is the anti-Klebanov-Strassler solution, and there is no regular solution whose D3-charge is negative in the IR and positive in the UV.

\end{abstract}

\maketitle


  {\bf Introduction.} A nonzero positive cosmological constant appears to be the most plausible cause for the observed accelerated expansion of our universe, and thus, in order to be a candidate for a theory of everything, string theory must contain low-energy de Sitter (dS) space solutions. On the other hand, the generic low-energy compactifications of string theory on six-dimensional manifolds with flux produces very large numbers of Anti de Sitter (AdS) vacua, but does not produce classical dS solutions with a cosmological constant smaller than the compactification scale.

To obtain phenomenologically-relevant dS solutions one needs to uplift the negative cosmological constant of AdS to a positive one, without disturbing the delicate balance needed to keep the compact dimensions stable, and the only known mechanism for doing this is to place objects with D-brane charge opposite to that of the background
(like anti-D3 branes \cite{Kachru:2003aw}) in regions of high redshift (or high warp factor) of the latter.
This ensures that the contribution of the anti-branes to the cosmological constant can be parametrically small, and implies that the many AdS low-energy flux compactifications can be uplifted to dS vacua, and hence string theory has a landscape of dS low-energy vacua.

The best-studied model for a highly-warped region of a flux compactification is the so-called Klebanov-Strassler warped deformed conifold (KS) solution \cite{Klebanov:2000hb}, and anti-D3 branes placed in this solution have been argued to be metastable \cite{Kachru:2002gs} and are the key ingredient in the KKLT mechanism for uplifting AdS vacua and producing a de Sitter landscape \cite{Kachru:2003aw}. The suitability of anti-D3 branes in KS throats for describing metastable vacua and for uplifting AdS to dS vacua has been recently put into question by the perturbative investigation of the backreaction of these anti-branes \cite{Bena:2009xk,Bena:2011wh}, which found that near the anti-branes the solution develops an unphysically-looking singularity, and hence anti-D3 branes in KS may not give an asymptotically-decaying small deformation of this solution.

The stakes raised by this investigation are very high. If the singularity is not an artifact of perturbation theory (as suggested by \cite{Dymarsky:2011pm}), and if moreover it cannot be resolved in string theory, this implies that all solutions with anti-D3 branes in backgrounds with D3 charge dissolved in fluxes are unphysical. This would invalidate the KKLT mechanism for uplifting AdS vacua to dS ones, and imply that string theory does not have a landscape of vacua with a small positive cosmological constant.

\begin{figure}[h]
\begin{center}
\includegraphics[scale=0.5]{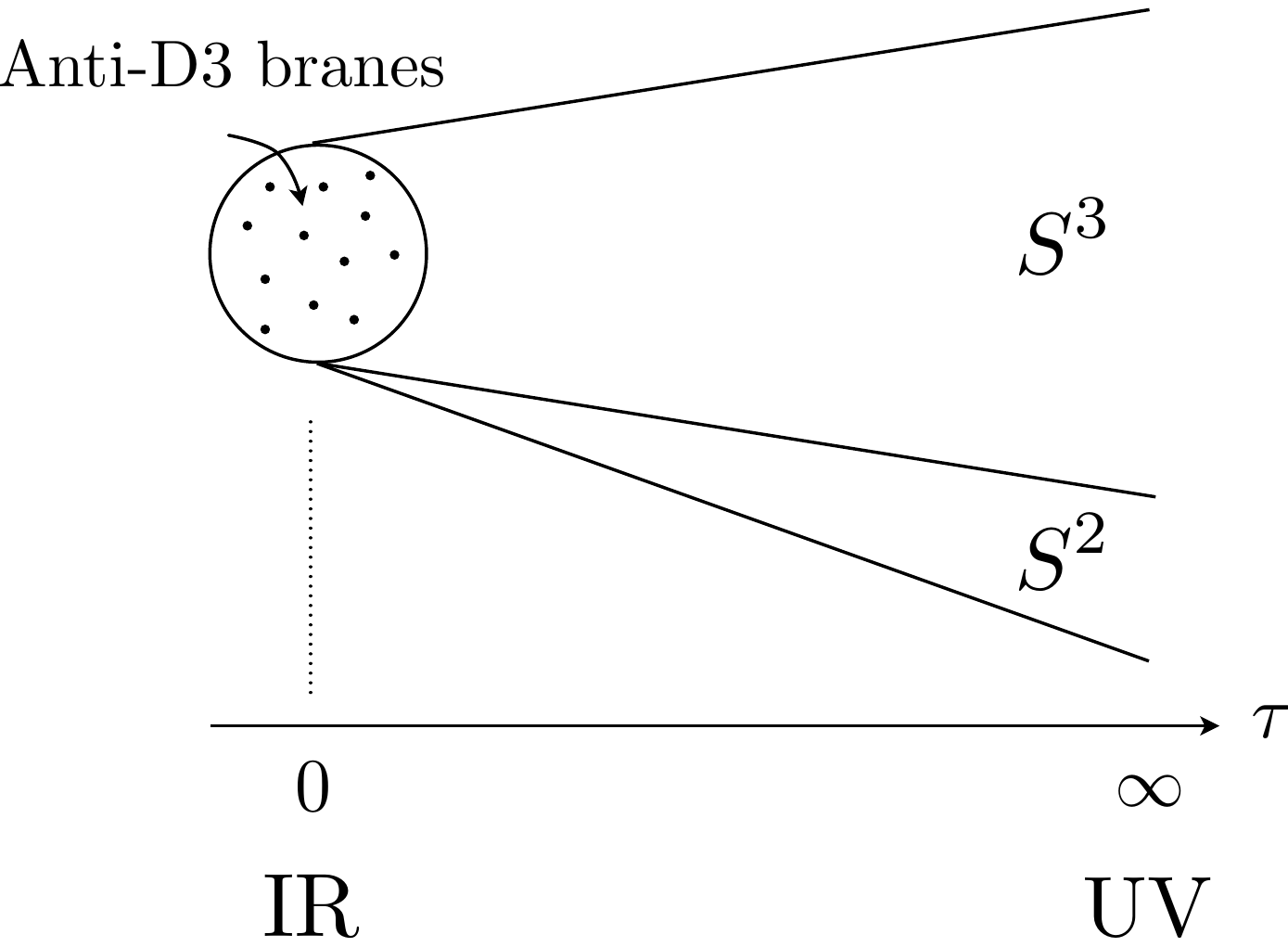}
\caption{The deformed conifold with anti-D3 branes smeared over the
  three-sphere at the tip ($\tau=0$). This setup is a concrete model
  of a warped throat in flux compactification, of the kind used to
  uplift AdS vacua to dS ones.}
\label{Setup}
\end{center}
\end{figure}

The purpose of this letter is to demonstrate that there exists \emph{no} fully-backreacted singularity-free solution describing smeared anti-D3 branes in a warped deformed conifold (KS) background with positive D3 charge dissolved in fluxes. Furthermore, the only fully-backreacted {\em regular} solution with anti-D3 branes in the infrared has anti-D3 charge dissolved in fluxes, and hence it is just the supersymmetric KS solution with a different charge orientation (which we will refer to as {\em anti-KS}). This was first conjectured in \cite{Bena:2009xk} (based on an analogy with the brane-bending calculation of \cite{Bena:2006rg}) and our results confirm this conjecture. The setup is shown on Figure \ref{Setup}.

To make such a statement one may naively try to construct the
fully-backreacted anti-D3 solution by solving analytically or
numerically the underlying 8 nonlinear coupled second-order
differential equations \cite{Papadopoulos:2000gj}, but this is not
necessary. We believe there exist at least {\it three} ways to demonstrate that
imposing regularity near the anti D3-branes cannot give a solution
with positive D3 charge at infinity, and in this note we present the three proofs: 

\smallskip

\begin{inparaenum}

   \item  We solve brute-force the equations in a Taylor expansion
     around the infrared. Setting to zero all the coefficients that
     give singular metric and 3-form fluxes, we found that the full
     solution up to order $\tau^{10}$ (where $\tau$ is the radial
     coordinate away from the KS tip) has three independent parameters
     all of which, as we will show below, are singular in the
     ultraviolet. The only \emph{regular} solution is hence the BPS
     anti-KS solution with anti-D3 branes. \smallskip

   \item  We explore the boundary conditions for the fields and their
     derivatives in the infrared, and show that if one imposes
     singularity-free boundary conditions the right hand sides of some
     of the equations are zero at all orders in perturbation
     theory. The remaining equations only have UV-singular solutions,
     and the only possible regular solution is the supersymmetric
     one. \smallskip

   \item  A more elegant way to prove that there is no regular solution whose D3-brane charge changes sign from IR to UV is to find a topological argument similar to that of \cite{Blaback:2011nz}. We present an argument along these lines. This argument may be generalizable to the case of \emph{localized} anti-D3 branes.

 \end{inparaenum}

\smallskip

  {\bf The setup.}
As argued in \cite{Bena:2009xk}, the Ansatz for the solution describing smeared D3 and anti-D3 branes in the KS solution is \cite{Papadopoulos:2000gj}:
\begin{eqnarray}
    d s_{10}^2 &=& e^{2\, A+2\, p-x}\, d s_{1,3}^2 + e^{-6\, p-x}\, \left( d\tau^2 + g_5^2 \right)   \nonumber \\
    \quad &&
    + e^{x+y}\, \left( g_1^2 + g_2^2 \right) + e^{x-y}\, \left( g_3^2 + g_4^2 \right) \, \label{PTmetric} \\
 H_3 &=& \frac{1}{2} \left( k - f \right) \, g_5 \wedge \left( g_1 \wedge g_3+ g_2 \wedge g_4 \right) \nonumber  \\
    \quad &&
        + \, d\tau \wedge \left( \dot{f} \, g_1 \wedge g_2 + \dot{k} \, g_3 \wedge g_4 \right)  \nonumber  \\
 F_3 &=& F \, g_1 \wedge g_2 \wedge g_5 + \left( 2 P - F \right) \, g_3\wedge g_4 \wedge g_5
    \nonumber \\
    \quad &&
     + \, \dot{F} \, d \tau \wedge \left( g_1 \wedge g_3 + g_2 \wedge g_4 \right) \,  \label{PTfluxes} \\
 F_5 &=& {\cal F}_5 + * {\cal F}_5 \, ,
    \quad {\cal F}_5 = K \, g_1 \wedge g_2 \wedge g_3 \wedge g_4 \wedge g_5 \, , \nonumber
\end{eqnarray}
with
\begin{equation}
\label{K}
   K =  - \frac{\pi}{4} Q + (2P - F) f + k F \, ,
\end{equation}
where all the functions depend only on the radial variable $\tau$ and
the angular forms $g_i$ are defined in \cite{Klebanov:2000hb}.
The constant $P$ is proportional to the $5$-brane
flux of the KS solution and $Q$ is the number
of (anti) D3 branes.

In order to handle the second-order equations of motion for the
scalars of the PT Ansatz, we found crucial to define particular
combinations of fields, inspired by the GKP~\cite{Giddings:2001yu} notations. The
warp factor $e^{4A+4p-2x}$ and the five-form
flux $K \text{vol}_5$ are combined into scalar modes $\xi^{\pm}_1$, defined as
\begin{equation}
  \xi_1^{\pm}=  - e^{4(p + A)} \left( \dot{x} - 2 \dot{p} - 2 \dot{A}
    \mp \frac{1}{2} e^{-2 x} K  \right) \, .
\end{equation}
These modes have a clear physical interpretation: they parametrize
the force on probe D3 and anti-D3 branes in a given solution:
\begin{equation} \label{forceD3}
F_{D3} = - 2 e^{-2x} \xi_1^{+} \ , \quad F_{\overline{D3}} = - 2 e^{-2x}  \xi_1^{-}\, .
\end{equation}
We also introduce ISD and IASD three--form
fluxes:
\begin{equation}
G_{\pm} = (\star_6 \pm i ) G_3 \, ,
\end{equation}
with $\star_6$ the six-dimensional Hodge star and $G_3 = F_3 + i
e^{-\phi}H_3$. The scalar components of $G_{\pm}$ will be called $\xi_f^{\pm}$, $\xi_k^{\pm}$ and
$\xi_F^{\pm}$. This notation follows from the fact that these modes
are the conjugate momenta to the fields $f, k$ and $F$
in~\eqref{PTfluxes}, in the reduced one-dimensional system that
describes the dynamics of the 8 scalar functions (seven in \eqref{PTmetric} and \eqref{PTfluxes} plus the dilaton $\phi$).

Supersymmetry imposes either that $G_{-} = F_{D3} = 0$ or $G_{+} =F_{\overline{D3}}  = 0$,
depending on which supersymmetries are preserved.
We will refer to the solutions with ISD and IASD fluxes as KS
and \emph{anti-KS} respectively. With this notation the KS solution
has $\xi_a^{+}=0$, $a=1,f,k,F$, while for the anti-KS solution $\xi^{-}_a = 0$.
A crucial fact is that the equations of motion for the scalars $\xi^{\pm}_a$ are just first-order
ODEs. For the $\xi_a^{-}$ modes we find:
\begin{align}
\label{dx1-eq}
    & \dot{\xi}^{-}_1 + K e^{-2 x} \xi^-_1  = \\
    & \quad
      4 e^{2 x - 4(p + A)}
        \left[ e^{\phi+2 y}  (\xi^-_f)^2 + e^{\phi- 2 y} (\xi^-_k)^2
        + \frac{1}{2} e^{-\phi} (\xi^-_F)^2 \right]  \nonumber
\end{align}
and
\begin{eqnarray}
\label{dxfkF-eq}
    \dot{\xi}^-_f &=& \frac{1}{2} e^{-2 x} (2 P - F) \xi^-_1 + \frac{1}{2} e^{-\phi} \xi^-_F    \nonumber \\
    \dot{\xi}^-_k &=& \frac{1}{2} e^{-2 x} F \xi^-_1 - \frac{1}{2} e^{-\phi} \xi^-_F   \\
    \dot{\xi}^-_F &=& \frac{1}{2} e^{-2 x} (k - f) {\xi}^-_1 +
                    e^{\phi} \left( e^{2 y} \xi^-_f - e^{-2 y} \xi^-_k \right)   \, .   \nonumber
\end{eqnarray}
Remarkably, these are the only equations that we will need in this letter.
One can define additional scalars $\xi^{\pm}_b$ which are the conjugate
momenta for the four additional
modes $x,y,p,\phi$, in such a way that
the BPS KS solution with $Q$ mobile D3-branes has all the
$\xi^{+}$ modes equal to zero. The 8
integration constants  of the BPS system $\xi^{+} = 0$ are fixed as
follows:

\begin{inparaenum}
\item The zero-energy condition of the effective Lagrangian fixes the $\tau$-redefinition gauge
        freedom and is automatically solved when $\xi_a=0$, but the constant shift
        $\tau \to \tau + \tau_0$ still remains unfixed, and so
        $\tau_0$ appears as a ``trivial" integration constant.
 \item The conifold deformation parameter $\epsilon$ and the constant dilaton $e^{\phi_0}$ give
        two other free parameters.
  \item An additional parameter renders the conifold metric singular
    in the IR \cite{Candelas:1989js} and has to be discarded.
  \item The three equations for the flux functions $f$, $k$ and $F$
    appear to have three free parameters \cite{Kuperstein:2003yt}. One
    gives singular BPS fluxes in the IR, the second gives a $(0,3)$
    complex 3-form $G_3 \equiv F_3 + i e^{-\phi} H_3$ that is singular
    in the UV, and the third corresponds to a $B$-field gauge
    transformation $(f,k) \to (f + c, k + c)$ that can be absorbed in
    the redefinition of $Q$.
  \item The warp function $h \equiv e^{-4(p + A) + 2 x}$ can only be determined up to a constant,
        which is fixed requiring that $h$ vanishes at infinity.
\end{inparaenum}

\smallskip
To summarize, the KS solution with $Q$ mobile D3-branes and the free parameters $\epsilon$ and $e^{\phi_0}$
is the only (IR and UV) regular solution with $\xi^{+}_a = 0$, where by
IR-regular we denote a solution whose only singularities are those
coming from D-branes.


\smallskip

 {\bf The boundary conditions for anti-D3-branes.}
The main goal of this letter is to show that there is no IR-regular solution
with smeared anti-D3 branes ($Q<0$, hence $K>0$) at the tip of the
conifold and with KS asymptotics ($K<0$) in the UV.
Starting with a singularity-free anti-brane solution in the IR, one necessarily ends up with an anti-KS solution in the UV.
Moreover, we will prove that the \emph{only} regular solution with
$|Q|$ anti-D3 branes is the anti-KS flip of the solution with $Q$
mobile branes we reviewed above.

To obtain IR-regular solutions we require that:
\smallskip

\begin{inparaenum}[\textbullet]
  \item the $6d$ conifold metric has the tip structure of the KS solution:
      the 2-sphere shrinks smoothly
        at \mbox{$\tau=0$} and the 3-sphere has finite size. \smallskip

  \item The warp factor comes from $|Q|$ anti-D3 branes smeared on the 3-sphere, and hence goes like  \mbox{$ h \sim |Q|/\tau$}.
        As a result the Taylor expansions of the functions $x$, $p$, $A$ and $y$ start
        with the same logarithmic \emph{and} constant terms as in the KS solution with mobile branes and can differ
        only by linear (and higher) terms. The constant term in $A$ cannot be fixed by the regularity condition,
        since it corresponds to the conifold deformation parameter
        $\epsilon$. \smallskip

  \item There is no singularity in the three-form fluxes; their energy densities, $H_3^2$ and $F_3^2$,
        do not diverge at  $\tau=0$. Hence, the Taylor expansions of the functions $f$, $k$ and  $F$
        start from $\tau^3$, $\tau$ and $\tau^2$ terms respectively,
        exactly like in the KS background. To be more precise, in a
        solution with branes at the tip the functions $f$, $k$ and $F$
        can also start with non-integer powers ($\tau^{9/4}$,
        $\tau^{1/4}$ and $\tau^{5/4}$), but one can show that the
        logarithmic terms in the metric imply that the IR expansion of
        the solution only has integer powers of $\tau$. \smallskip

  \item The dilaton is finite at $\tau=0$. \smallskip
\end{inparaenum}

It is important to stress that we \emph{do not} impose any kind of anti-KS
IR boundary conditions for the 3-form fluxes, and a-priori the 3-form can be either ISD or IASD (or have both components).
On the other hand, we do require the singularities in the warp factor and the five-form flux to correspond to objects that exist in string theory.

These observations are helpful to determine the possible leading-order behaviors of the $\xi^+_a$'s and $\xi^-_a$ (for our argument we mostly need the latter).
Let us denote by $n_a$ the lowest \emph{possible} leading orders of the fields $\xi^-_a$.
For small $\tau$, the metric regularity conditions imply that the
functions $e^{2x}$ and $e^{4 (p + A)}$ go like $\tau$ and $\tau^2$
respectively.
From the explicit definitions of the $\xi_a^-$ modes we find:
\begin{equation}
\label{powers}
    \left( n_1, n_f, n_k, n_F \right) =  \left( 2, 1, 3, 2\right) \, .
\end{equation}

\smallskip

 {\bf The IR obstruction.}
Our goal is to show that when solving the equation of motions
(\ref{dx1-eq}), (\ref{dxfkF-eq}) for $\xi^-_1$, $\xi^-_f$, $\xi^-_k$
and $\xi^-_F$ in the IR (small $\tau$) and imposing the IR regularity
conditions, one finds only trivial solutions for these functions. This
essentially means that the IASD conditions $\xi^-_f=\xi^-_k=\xi^-_F =
0$ will be satisfied all the way to the UV and not only at
$\tau=0$. To prove this, a simple counting argument is sufficient, as
we will now prove.


Let us assume that $\xi^-_F$ and $\xi^-_1$ start from $\tau^n$ and
$\tau^{n+l}$ for some $n \geqslant 2$.
 We treat separately the two
possibilities:\smallskip
\begin{inparaenum}

  \item $l > -1$. Recalling that $e^y \approx \frac{\tau}{2} + \ldots$, we can see from a simple power analysis that the $\xi^-_1$ term is subleading both in the $\xi^-_k$ and $\xi^-_F$ equations in (\ref{dxfkF-eq}).
      In the latter equation the $\xi^-_f$ term is also subleading. We arrive at the set of two simple equations near $\tau=0$: $\dot{\xi}^-_k \approx - \frac{1}{2} e^{-\phi_0} \xi^-_F$ and
      $e^{- \phi_0} \dot{\xi}^-_F \approx - 4 \tau^{-2} \xi^-_k$. They have only two solutions,
      $\xi^-_F \sim \tau^{-2}$ and $\xi^-_F \sim \tau$ and
      both fall short of the regularity conditions
      (\ref{powers}). Remarkably, in  showing that the system has no
      regular solution we have not used (\ref{dx1-eq}).\smallskip

  \item $l \leqslant -1$.  Now the right hand side of (\ref{dx1-eq}) is certainly negligible with respect
        to the left hand side.
       This means that we have $\dot{\xi}^-_1 + \tau^{-1} \cdot \xi^-_1 \approx 0$  for small $\tau$ leading to the singular solution $\xi^-_1 \sim \tau^{-1}$. \smallskip
\end{inparaenum}

For non-integer powers, the argument above can be straightforwardly extended, and the two regimes of parameters corresponding to those above are  $l > -1/4$ and $l \leqslant -1/4$.
\smallskip

To conclude, we see that regularity in the IR implies that the functions $\xi^-_1$, $\xi^-_f$, $\xi^-_k$
and $\xi^-_F$ vanish identically. Consequently, the solution will remain IASD for any value of $\tau$, and the force on probe anti-D3 branes will remain identically zero.

A second way to see that a solution with negative D3 charge in the
infrared remains anti-KS all the way to the UV is to solve directly
the second-order equations for the scalar in the PT Ansatz in a power
expansion around the origin. Upon eliminating all the singular modes,
we find that to order $\tau^{10}$ the space of solutions is
parameterized by  \emph{three} constants. None of these constants
breaks the IASD condition, which confirms the results of the previous
section.

One can also use the fact that $\xi^-_1$, $\xi^-_f$, $\xi^-_k$, $\xi^-_F$ are necessarily zero to identify the three IR modes we find, and to show that they correspond actually to
UV singular solutions:\smallskip
\begin{inparaenum}

  \item
 Plugging $\xi^{-}_{1,f,k,F}=0$ into the remaining equations of motion,
it is easy to show that there exists an IR regular but UV divergent
one parameter family of solution. \smallskip

  \item A second mode is the $(3,0)$-form solution of the
    superpotential $\xi^{-}_a=0$ equations, which breaks
    supersymmetry and diverges in the UV (see
    \cite{Kuperstein:2003yt}).\smallskip

  \item A third ``superpotential mode" is related to the shift of the warp function
        and, following our previous discussion, has to be excluded.
\end{inparaenum}

\smallskip
Summarizing, we see that the only solution with smeared \emph{anti}-D3
branes at the KS tip that is regular both in the UV and in the IR is
the \emph{anti}-KS solution. Stated differently, the only way to
obtain a sensible supersymmetry-breaking solution corresponding to the
backreaction of smeared anti-D3's is to allow for IR
singularities in the energy densities of the 3-form fluxes.

\smallskip
\smallskip

\smallskip
 {\bf The global obstruction.}
We can also present a ``global" argument why the  functions $\xi^{-}_1$, $\xi^{-}_f$, $\xi^{-}_k$ and $\xi^{-}_F$ have to vanish in a regular solution, without focusing on their Taylor expansions. The proof for the remaining four functions proceeds precisely as above.

Our key observation is that the flux functions $f(\tau)$, $k(\tau)$ and $F(\tau)$ appear only in equations (\ref{dx1-eq}) and
(\ref{dxfkF-eq}). None of the remaining $\dot{\xi}^{-}_a$ equations has any flux function in it. Next, the equations in
(\ref{dxfkF-eq}) might be derived from the following \emph{reduced} Lagrangian:
\begin{align}
    \mathcal{L}_{\rm fluxes} &= 4 e^{2 x - 4(p + A)+\phi}
        \Big[e^{2 y}  (\xi^{-}_f)^2 + e^{- 2 y} (\xi^{-}_k)^2
        \nonumber \\
          & \qquad     + \frac{1}{2} e^{-2\phi} (\xi^{-}_F)^2 \Big]
        + e^{- 4(p + A)} (\xi^{-}_1)^2  \, . \label{Lreduced}
\end{align}
We treat $\mathcal{L}_{\rm fluxes}$ as the effective Lagrangian \emph{only} for the  fields $f(\tau)$, $k(\tau)$ and $F(\tau)$ with the remaining five fields being free but subject to the proper boundary conditions ensuring the IR regularity. This means that the first three terms in (\ref{Lreduced}) are kinetic terms, while the last one is a potential term.
Recall also that the $\xi^{-}$'s are first order in the derivatives of $\phi$'s and so the Lagrangian is of the second order, precisely as it should be.

This Lagrangian has a remarkable property: it is strictly non-negative and vanishes only for  $\xi^{-}_1, \xi^{-}_f, \xi^{-}_k, \xi^{-}_F = 0$. In other words, the \emph{global} minimum of (\ref{Lreduced}) corresponds to the IASD solution. The only way to arrive at a different solution, which describes only a \emph{local} minimum of the action, is to impose boundary conditions (either in the IR or in the UV) that are at odds with the IASD solution $\xi^{-}_{f,k,F,1}=0$.

The regularity requirement, however, constrains all the three flux functions and their conjugate momenta $\xi^{-}_{f,k,F}$ in the IR. Indeed, we saw that both $(f, k, F)$ and $( \xi^{-}_f, \xi^{-}_k, \xi^{-}_F )$ have to vanish at $\tau=0$ for a regular solution. Similarly $\xi^{-}_1=0$ in the IR, thus the IR boundary conditions following solely from the regularity are consistent with the ``trivial" IASD solution. We conclude again that requiring regularity forces upon us the anti-KS solution.

Note, however, that the equations derived from (\ref{Lreduced}) are singular in the IR, and so our arguments should be taken very cautiously. At the same time, this approach may prove efficient for the localized anti D3 branes, where one cannot use the Taylor expansion argument.

\smallskip
 {\bf Conclusions.}
We presented a detailed analysis of the nonlinear backreaction of
smeared anti-D3 branes on the KS geometry. In the near-brane (IR) region we
impose boundary conditions coming from the presence of a smeared
source: singular warp factor and commensurate five-form flux, while in
the UV region we require absence of highly divergent modes. We
showed that with these assumptions there is a unique solution of the
equations of motion, namely the supersymmetric anti-KS solution.

We have thus proven that any supersymmetry-breaking solution associated to the
backreaction of smeared anti-D3 branes on the KS geometry has
singularities in the IR region not directly associated with the anti-D3 branes themselves. Moreover, these singularities appear in the
three-form fluxes, confirming the linearized analysis
of~\cite{Bena:2009xk}.

A singularity in a supergravity solution does not necessarily mean that this solution should be automatically discarded. However, physically acceptable singularities are those which are resolved in the full string theory (for example, the singularity in the supergravity solution for a brane is resolved by the open strings on the brane, other singularities are resolved by brane polarization \cite{Myers:1999ps,Polchinski:2000uf} or geometric transitions). Here it is not at all clear that there is any mechanism capable of explaining the present singularities. If there is no such mechanism, then the singularity in the supergravity background is telling us that  there is no (meta)stable anti-D3 brane solution, and the whole system of branes with charge opposite to that of the background is unstable. This would invalidate the anti-brane AdS to dS uplifting mechanism, and therefore most of the String Theory de Sitter landscape.

\noindent{\bf Acknowledgements:} We would like to thank Thomas van Riet for stimulating discussions. This work was supported in part by the ANR grant 08-JCJC-0001-0 and the ERC Starting Grants 240210 - String-QCD-BH and 259133 -- ObservableString.

\begingroup\raggedright\endgroup

\end{document}